# Probing the structure and dynamics of the actinides from U through Lr using Regge-pole analysis: A Mini Review

Zineb Felfli
Department of Physics and Astronomy, Georgia State University, Atlanta, Georgia 30303, USA

Zoe Landers and Alfred Z. Msezane[*]
Department of Physics and Center for Theoretical Studies of Physical Systems, Clark Atlanta University, Atlanta, Georgia 30314, USA

**Abstract**.
The structure and dynamics of the actinide atoms from U to Lr is probed through the electron elastic scattering total cross sections (TCSs) calculated using Regge pole analysis. The crucial Regge trajectories probe electron attachment at the fundamental level near threshold leading to stable ground, metastable and excited negative-ion formation as resonances as well as Ramsauer-Townsend (R-T) minima and shape resonances (SRs). A polarization-induced metastable TCS with a deep R-T minimum near threshold first appears in the Pu TCSs, marking the first break in periodicity in the actinides while in the Cf TCSs the R-T minimum manifests as a SR very close to threshold, demonstrating the second break. These results are consistent with observations. Moreover, the highest excited states TCSs of U and Pu exhibit the strongest fullerene molecular behavior among the actinides, implying their multi-functionalization, including nanocatalysis through their doubly- charged anions. Electron affinities are also extracted from the TCSs.

**Keywords:** Regge poles, Ramsauer-Townsend minima, anionic binding energies, polarization-induced metastable, actinide atoms, electron affinities, elastic cross-sections

## 1. Introduction

The actinide elements are highly radioactive; this renders them vital for both scientific research and practical applications, ranging from cancer diagnostics and treatment to renewable energy technologies and long-lived nuclear batteries for deep space exploration. The U and Pu elements are particularly essential in the fields of nuclear physics and chemistry. Pu, U and Th provide fuel in nuclear reactors while the former two are vital constituents of nuclear weapons and are also used in radiotherapy for targeted cancer treatment. Research in actinides advances to understand nuclear reactions and properties of heavy elements. For instance, at the Berkeley Lab researchers continue to better understand how Bk and Cf could accelerate new applications in medicine, energy and technologies. To this end, Bk has been utilized as a target nuclide in the synthesis of heavier transuranic and transactinide elements while Cf, a very strong neutron emitter, is used to identify gold and silver ores as well as to detect metal fatigue and stress in airplanes.

Unfortunately, the actinides' radioactive nature also renders them very challenging to handle experimentally [1]. Consequently, the important electron affinities (EAs) of only Th [2] and U [3,4] atoms have been measured thus far. Indeed, the EA provides a stringent test of theoretical calculations when the calculated EAs are compared with those obtained from reliable measurements. For instance, the measured EAs of Au [5-7], the highly radioactive At [8] and the fullerene molecular $C_{60}$ [9-11] correspond to the Regge pole-calculated negative ion binding energies (BEs) when the incident electron is attached in the ground state of the formed negative ion during the collision. The Regge pole-calculated negative ion BEs for Au, At, $C_{60}$, and Eu [1, 12] as well as the fullerene molecules $C_{20}$ through $C_{92}$ [13, 14] are in outstanding agreement with the measured EAs. And very sophisticated theoretical EAs of Au [15-17], the radioactive At [18-21] and Eu [22, 23] agree very well with the Regge pole-calculated BEs. This gives great credence to the ability of the Regge pole approach to obtain reliable EAs of multielectron systems without any assistance from either experiment or other theoretical methods.

In this paper we have selected the actinide atoms U, Np, Pu, Cm, Bk, Cf, No and Lr to illustrate the effective probing of their electronic structure and dynamics using low-energy electron impact. The paper has been motivated by the following:

1) The experiments [24] and [25] observed the first break in periodicity in the actinide elements occurring between the Pu and Am elements, while the second break appears between the Bk and Cf elements [26]. The great significance of the investigation [26] lies in that the experiment used only a single nanogram of the highly radioactive Bk and Cf atoms to characterize their structure and dynamics. This together with the recent EA measurement of the highly radioactive At [8] promise far-reaching consequences in measurements involving the radioactive elements; it could speed up the long overdue measurements of the EAs of many of the actinide elements.
2) The existing sophisticated theoretical calculations incorporating relativistic effects obtained the EAs of most actinides that differ significantly in their predictions. And importantly, for many of the actinide elements the predicted EAs [27] are mostly negative and differ significantly from those of [28]. For Lr there are other EAs [29, 30] calculated using elaborate theoretical methods that are normally not exhibited in comparisons.
3) We would like to identify actinide atoms that exhibit fullerene molecular behavior [13] for novel use in catalyzing water oxidation to peroxide through their singly or doubly charged negative ions [31, 32]. The fundamental mechanism underlying negative-ion catalysis involves bond-strength breaking/reformation in the transition state [33]. And doubly charged atomic/molecular anions have been proposed as novel dynamic tunable catalysts, with $Si^{2-}$, $Pu^{2-}$, $Pa^{2-}$ and $Sn^{2-}$ being the best catalysts, the radioactive elements usher in new application opportunities. Indeed, exploiting their radioactive nature, tunability and wide applications, ranging from water purification to biocompatibility would certainly revolutionize the field of nanocatalysis.

**2. Method of calculation**

The stringent conditions imposed on the Regge trajectories and the sensitivity of the stable negative ion-formation to the polarization interaction potential in low-energy electron collisions with multielectron atoms and fullerene molecules have been exploited and used to extract rigorous negative ion BEs from the calculated TCSs [34]. This has proved to be the great strength and asset of the Regge pole method to low-energy electron scattering. Within the framework of the CAM description of scattering involving Regge poles we have thus probed with unprecedented success through the TCSs calculations negative ion formation in low-energy electron collisions with multielectron atoms and fullerene molecules, leading to stable ground, metastable and excited negative ion formation.

Regge poles, have been studied extensively close to a century in such diverse fields as atomic and molecular theory (see the relevant reviews [35] and [36]) and high energy physics [37]. In [35] applications of the Regge pole analysis focused on the interpretation of resonance structures observed in elastic, inelastic and reactive angular distributions as well as in total elastic cross-sections of atom–atom and electron–atom collisions. Also, the need for accurate Regge poles positions and residues for potential scattering models, required for TCSs calculations, has led to the development of accurate methods for calculating these parameters (see, for example [38], [39], [40] and references therein).

The details of the calculation of the TCSs describing the collision between low-energy electrons and multielectron systems using the Regge pole method are found in [41], [42]. As demonstrated in Figures 1(a) and 1(b) of the recent paper [34] the vital Regge trajectories, also important quantities calculated in CAM theories, yield through the TCSs the rigorous BEs of the negative ions formed during the collision. Significantly, in case of electron collision with multielectron systems (atoms and fullerene molecules) the source of the bound states generating the Regge trajectories is the attractive Coulomb well it experiences near the nucleus. The addition of the centrifugal term to the well "squeezes" these states into the continuum [36, 42]. Consequently, a bound state crossing the threshold energy E = 0 in this region may become an excited state or a long-lived metastable state. As a result, the highest "bound state" formed during the collision is identified with the highest excited state, here labeled as EXT-1; the next excited state if it exists, is labelled as EXT-2, etc. Notably, the metastable states are labeled relative to the ground state, namely as MS-1, MS-2, beginning from the ground state, located at the absolute minimum of the ground state TCS.

Briefly, within the CAM description of scattering [35], Regge poles, singularities of the S-matrix, rigorously define resonances [43, 44]. In the physical sheets of the complex plane, they correspond to bound states [45] and those formed during low-energy electron elastic scattering become stable bound states [46]. In the near-threshold electron–atom/fullerene molecule collision resulting in stable negative ion formation as resonances, the TCSs calculation exploits the Regge pole method through the Mulholland formula [47]. This formula converts the notoriously very slowly convergent infinite discrete sum into a background integral plus the contribution from a few poles to the scattering process [41, 42]. Indeed, the method requires *no a priori* knowledge of the experimental or any other theoretical data as guidance; hence, its predictive nature. The Mulholland formula used here is of the form [41, 42] (atomic units are used throughout):

$$\sigma_{tot}(E) = 4\pi k^{-2} \int_0^\infty \text{Re}[1 - S(\lambda)]\lambda d\lambda$$
$$- 8\pi^2 k^{-2} \sum_n \text{Im} \frac{\lambda_n \rho_n}{1 + \exp(-2\pi i \lambda_n)} + I(E) \tag{1}$$

In Eq. (1) $S(\lambda)$ is the S-matrix, $k = \sqrt{2mE}$, $m = 1$ being the mass and E the impact energy, $\rho_n$ is the residue of the S-matrix at the $n^{th}$ pole, $\lambda_n$ and $I(E)$ contains the contributions from the integrals along the imaginary $\lambda$-axis ($\lambda$ is the complex angular momentum); its contribution has been demonstrated to be negligible [23].

As in [48], we consider the incident electron to interact with the complex heavy system without consideration of the complicated details of the electronic structure of the system itself. Thus, the robust Avdonina–Belov–Felfli potential [49], which embeds the vital core–polarization interaction is used:

$$U(r) = -\frac{Z}{r(1 + \alpha Z^{1/3} r)(1 + \beta Z^{2/3} r^2)} \tag{2}$$

In Equation (2) Z is the nuclear charge, $\alpha$ and $\beta$ are variation parameters. For small r, the potential describes Coulomb attraction between an electron and a nucleus, $U(r) \sim -Z/r$, the source of the bound states generating the crucial Regge trajectories, while at large distances it has the appropriate asymptotic behavior, *viz.* $\sim -1/(\alpha\beta r^4)$ and accounts properly for the polarization interaction at low energies.

The strength of this extensively studied potential, Equation (2) [38, 50, 51, 52] lies in that it has five turning points and four poles connected by four cuts in the complex plane. The presence of the powers of Z as coefficients of r and $r^2$ in Equation (2) ensures that spherical and non-spherical atoms and fullerenes are correctly treated. Small and large systems are also appropriately treated. The effective potential $V(r) = U(r) + \lambda(\lambda+1)/(2r^2)$ is considered here as a continuous function of the variables r and complex $\lambda$. Importantly, our choice of the potential, Equation (2) is adequate as long as our investigation is confined to the near-threshold energy regime, where the elastic TCS is less sensitive to short-range interactions and is determined mostly by the polarization tail.

Here the colliding partners (electron-atom/fullerene) are assumed to form a long-lived intermediate complex, manifesting as a sufficiently narrow resonance that rotates as it decays at zero scattering angle to preserve the total angular momentum. If the complex has a long angular life, namely Im λ(E) << 1, it will return to forward scattering many times. For the resonance to contribute to the TCS two vital resonance conditions must be satisfied: 1) Regge trajectory, namely Im λ(E) versus Re λ(E) stays close to the real axis and 2) Real part of the Regge pole is close to an integer, see Figures 1(a) and 1(b) of [34].

The numerical calculations of the TCSs, the pole positions and the residues are carried out by solving the Schrödinger equation as described in [42], see also [53]. The parameters "α" and "β" of the potential in Equation (2) are varied, and with the optimal value of $α = 0.2$, the β-parameter is further varied carefully until the dramatically sharp resonance appears in the TCS, see [34]. This is indicative of stable negative ion formation during the collision and the energy position matches with the measured EA of the atom/fullerene molecule, as seen in Table 1.

## 3. Results
### 3.1 Overview

To facilitate the understanding of the representative results in Figures 1 through 4, we first provide their overview. Figure 1 contrasts the standard Regge pole-calculated TCSs for atomic Au and fullerene molecular $C_{60}$, taken from our paper [54]. Figure 2 compares the Regge pole TCSs for U with those of Pu. In the Pu TCSs is observed the first appearance of the polarization-induced metastable TCS with the deep R-T minimum near threshold (brown curve); it is indicative of the first break in periodicity in the actinide series [24, 25]. And both the U and Pu TCSs exhibit significant fullerene molecular behavior through the green curves [13]. Figure 3 contrasts the Bk and Cf TCSs. Importantly, just as in the Pu TCSs the polarization-induced TCS with the deep R-T minimum near threshold still appears in the Bk TCSs. However, in Cf this TCS with the deep R-T minimum characterizing the Pu and Bk TCSs has flipped over to a SR very close to threshold. This demonstrates the second break in periodicity in the actinide series [26]. In Figure 4 the TCSs for No and Lr are compared. As with the Cf TCSs, both are characterized by the polarization-induced TCS with the pronounced SR located very close to threshold; these TCSs exhibit neither atomic nor fullerene behavior. Consequently, it is expected that the sophisticated theoretical methods currently employed to calculate the EAs of these atoms will not be severely handicapped.

### 3.2 Au and $C_{60}$ Total Cross Sections

In the context of this paper, the important curves in Figure 1 for both Au and $C_{60}$ are the ground and the excited states TCSs, represented by the red and the green curves, respectively. In both Au and $C_{60}$ the EAs are determined by the BEs of the negative ions formed in the ground states, viz. 2.26 eV and 2.66 eV, respectively. It is further noted that any one of the sharp peaks, representing negative-ion formation in the metastable and excited states could be identified with the EAs if an experiment so determines. Notably, in [55] it was cautioned that the isomers (sharp peaks in the metastable TCSs of the $C_{60}$) could be misidentified with its EA. Indeed, the Regge pole calculated EA of $C_{60}$ [14] is in outstanding agreement with the measured EAs [9,10,11], see the comparison in Table 1. More recently, the Regge pole-calculated ground states negative-ion BEs were found to agree generally excellently with the measured EAs of the fullerene molecules from $C_{20}$ through $C_{92}$ [13, 14] for the first time; there are hardly any theoretical EAs for many of the fullerene molecules.

Calculating the EA of Au has been a daunting task for a long time, despite the presence of high-quality measurements, see Table 1. An outstanding agreement has now been achieved between the measured EA of Au [5,6,7] and that calculated using relativistic couple cluster method with variational quantum electrodynamics [15]. This formidable to calculate EA value, achieved by very few indeed, is also in excellent agreement with the Regge pole-calculated BE value for Au, obtained in 2008 [56], see also Table 1 for comparisons. Since here we are dealing with the highly radioactive actinide atoms, it is appropriate to comment on the recently measured EA of the highly radioactive At [8], which employed the coupled cluster method. The EA [8] agreed excellently with the Regge pole-calculated ground state BE [34] and the EAs from various sophisticated theoretical calculations, including the multiconfiguration Dirac–

Hartree–Fock values [18 - 21], see also Table 1 for comparisons. These results give great credence to the predictive power of the Regge pole analysis to produce robust EAs without any assistance whatsoever.

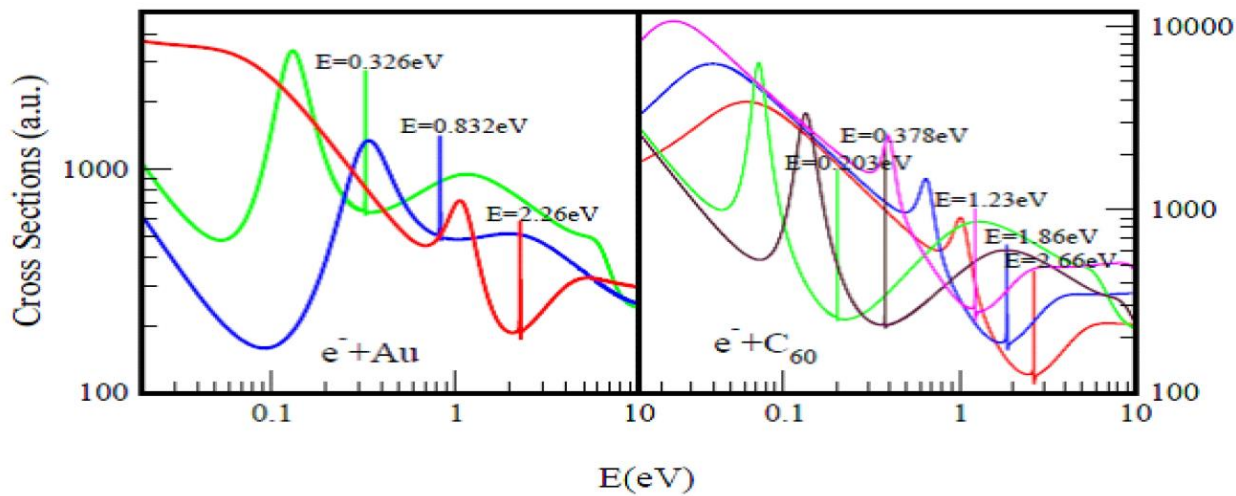

**Figure 1.** Standard total cross sections (TCSs) (a.u.) for electron elastic scattering from Au (left panel) and the fullerene molecule $C_{60}$ (right panel), taken from [54] are contrasted. For atomic Au, the red, blue and green curves represent TCSs for the ground, metastable and excited states, respectively. For the $C_{60}$ fullerene, the red, blue and pink curves represent TCSs for the ground and the metastable states, respectively, while the brown and the green curves denote TCSs for the excited states. The dramatically sharp resonances in both figures correspond to the $Au^-$ and $C_{60}^-$ negative ions formed during the collisions.

### 3.3 U and Pu Total Cross Sections

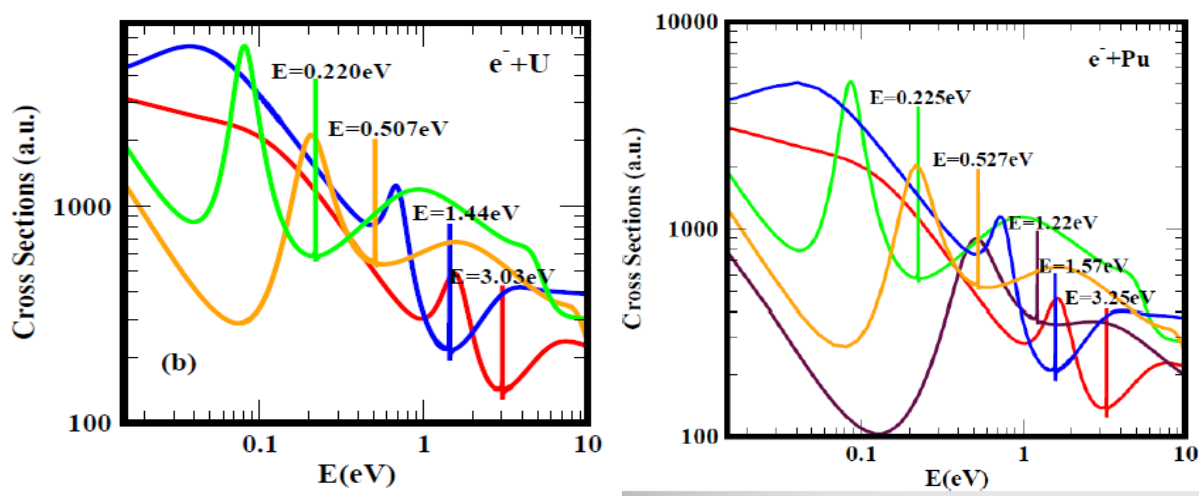

**Figure 2:** Total cross sections (a.u.) for U (left panel) and Pu (right panel). For U the red and blue curves represent TCSs for the ground and the metastable states, respectively, while for Pu the red, blue, and brown curves are TCSs for ground and metastable states, respectively. The green and the orange curves in U and Pu denote excited states TCSs. And the brown curve with the BE of 1.22eV in the Pu TCSs is the polarization-induced metastable TCS. The dramatically sharp resonances in the TCSs of both U and Pu correspond to $U^-$ and $Pu^-$ negative ion formation.

The U (left panel) and the Pu (right panel) TCSs shown in Figure 2, typifying those of the actinide atoms, are characterized generally by dramatically sharp resonances representing ground, metastable and excited states negative ion formation, SRs and R-T minima. The underlying physics has been discussed carefully in our paper [54]. The energy positions of the sharp resonances correspond to the rigorous anionic BEs of the formed negative ions during the electron collisions with the U and the Pu atoms. Like that of the Au the ground state TCSs of U and Pu fall off monotonically from threshold, while the metastable TCSs of both U and Pu show less pronounced SRs near threshold compared with those of the $C_{60}$ fullerene molecule. Also, the highest excited states TCSs of U and Pu exhibit strong fullerene behavior like those of the $C_{60}$ fullerene molecule [13]. Additionally, near threshold the metastable TCSs of the U and Pu atoms behave like those of the $C_{60}$ TCSs.

Importantly, the polarization-induced metastable TCS with the deep R-T minimum near threshold (brown curve in the Pu TCSs) first appears in the TCSs of the actinides in those of Pu. Indeed, this is a demonstration of the first break in periodicity in the actinide series, consistent with the observation [24, 25]. Moreover, the U and the Pu TCSs exhibit significant fullerene behavior [13] through their highest excited states TCSs. This renders them amenable to application in negative-ion catalysis [31, 32].

### 3.4 Bk and Cf Total Cross Sections

Recently, using only a single nanogram of the highly radioactive Bk and Cf elements, the experiment [26] discovered that a weak spin-orbit-coupling described the Bk structure, while that of Cf was characterized by a jj coupling scheme. It concluded that these observations strengthen Cf as a transitional element in the actinide series; indeed, this corresponds to the second break in periodicity in the actinide series [26]. The Bk and Cf TCSs presented in Figure 3 abound in very sharp resonances representing negative ion formation in the ground, metastable, and excited states as well as R-T minima and SRs. The energy positions of the dramatically sharp resonances in the TCSs yield the BEs of the formed negative ions during the collision. Notably, here the Bk

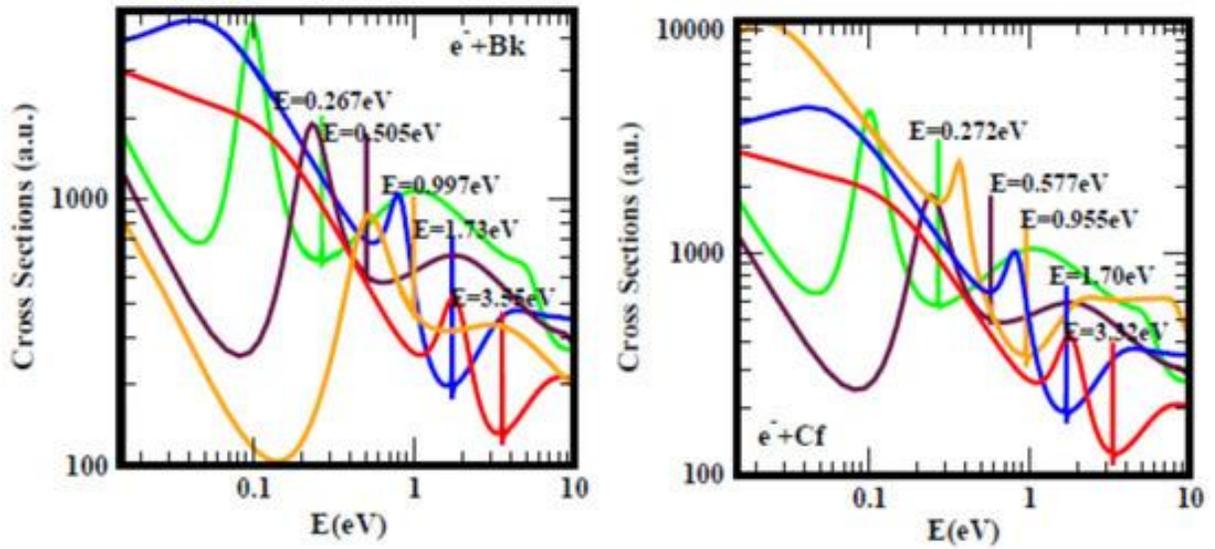

**Figure 3:** Total cross sections (a.u.) for atomic Bk (left panel) and Cf (right panel). For both Bk and Cf the red, blue, and orange curves represent TCSs for ground and metastable states, respectively, while the brown and the green curves correspond to excited state TCSs. The dramatically sharp resonances in the TCSs of both figures correspond to $Bk^-$ and $Cf^-$ negative ions formed during the collisions. Importantly, the flip over of the near-threshold deep R-T minimum from the Bk TCSs to a SR very close to threshold in the Cf TCSs occurs here.

TCSs are characterized by *inter alia* the polarization-induced metastable TCS with a deep R-T minimum near threshold, which first appeared in the Pu TCSs, see Figure 2. However, in the Cf TCSs this R-T minimum has flipped over to a SR very close to threshold. The paper [57] used the sensitivity of the R-T minimum and the SR in the polarization-induced metastable TCSs of Bk and Cf to validate the observations [26] for the first time. Incidentally, we could have used the ground state BEs in Table 1 to make our case. Since there are no measured EAs for these elements and the existing sophisticated theoretical EAs are ambiguous and difficult to understand, in Table 1 we have used the extracted BEs from the TCSs and compared them with the existing theoretical EAs.

Our tabulated BEs for the various actinide atoms in Table 1 are rigorous and can be used to guide both measurements and calculations. It is noted here that there are many sharp peaks in the TCSs representing ground, metastable and excited states negative ion formation during the collision, see the various Figures. Their positions correspond to the BEs of the formed negative ions during the electron collisions.

### 3.5 No and Lr Total Cross Sections

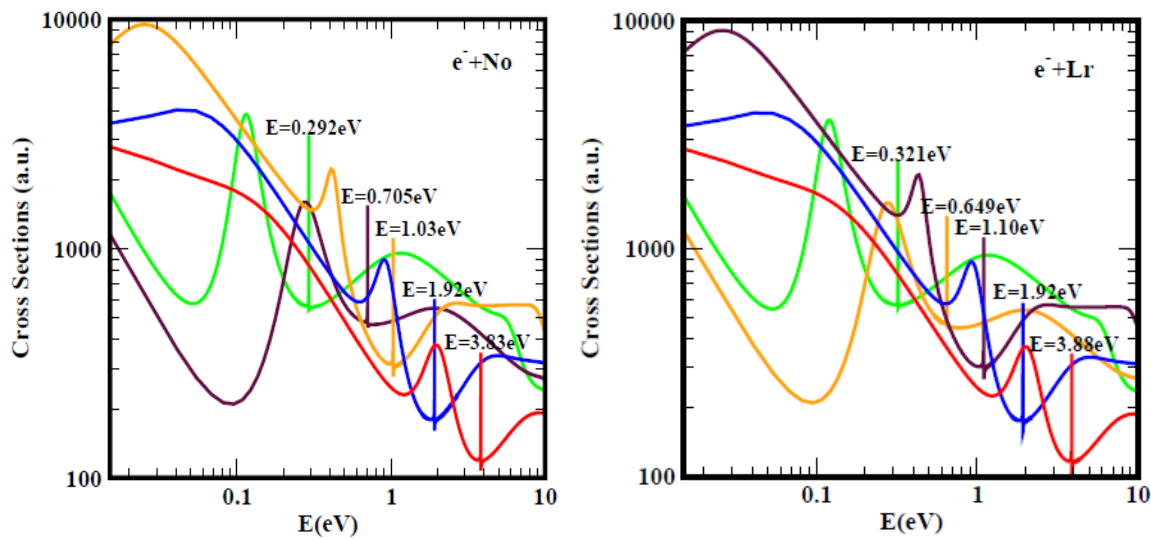

**Figure 4:** Total cross sections (a.u.) for atomic No (left panel) and Lr (right panel). For No the red, blue, and orange curves represent TCSs for ground and metastable states, respectively, while the brown and the green curves correspond to excited states TCSs. For Lr the red, blue, and brown curves denote TCSs for ground and metastable states, respectively, while the orange and the green curves correspond to excited states TCSs. The dramatically sharp resonances in the TCSs of both figures represent No⁻ and Lr⁻ negative ions formed during the collisions.

Figure 4 presents the TCSs of the No and Lr atoms; the location of these atoms near or at the end of the actinide series has motivated their selection. Consequently, their electronic structures are less complicated compared to those of Figures 2 and 3. Although there are no measured EAs for these actinides, nevertheless sophisticated relativistic theoretical EAs are available for Lr other than those of [27] and [28] to compare with. The TCSs of No and Lr resemble those of Figures 2 and 3. However, subtle differences are noted. The polarization-induced metastable TCS with the pronounced SR very close to threshold characterizes both the No and Lr TCSs; it first appeared in the Cf TCSs. Additionally, the highest excited states TCSs of both No and Lr exhibit neither atomic nor fullerene molecular behavior [13] and the very sharp resonances in their TCSs represent the BEs of the formed negative ions during the collisions. These BEs together with those from Figures 1 through 3 are summarized in Table 1, where they are compared with the available EAs. The inclusion of the EAs of Au, At, Eu and $C_{60}$ should help in guiding and interpreting the measurements and calculations of the EAs of the actinides.

The experimental determination with high precision of the BE of the least-bound electron in No, corresponding to the first ionization potential [58], and the measurements of the first ionization potentials of Fm, Md, No, and Lr [59], both guided by highly sophisticated theoretical calculations, promise the determination of the long overdue EAs of these highly radioactive elements. The Lr is the last element in the actinide series and its EA has not been measured yet, but it has been calculated [27, 28, 29, 30]. Indeed, for Lr, disagreements notwithstanding, all the theoretical EAs are positive, contrary to those of [27]. For instance, our EXT-1 BE value of 0.321eV is very close to the 0.310eV of [29] and 0.295eV of [28] as well as the Abs (-0.313) eV of [27]. This supports our remarks about the simpler structure of the Lr atom, minimizing the tremendous uncertainties facing theoretical calculations of the EAs of the more complex structurally actinide atoms. Do these data imply that the EA of the Lr atom corresponds to the Regge pole calculated BE of the highest excited state? Clearly, probing the structure and dynamics of the actinide atoms through Regge pole analysis yields unambiguous and reliable negative ion BEs.

### 3.6 Understanding the observed breaks in periodicity in the actinide atoms

In Figure 2 the polarization-induced metastable TCS, brown curve with the deep R-T minimum near threshold and a sharp resonance energy of 1.22eV, first appears in the Pu TCSs while those of U show no such a cross section. It has been attributed to size effects and orbital collapse [54], impacting the polarization interaction significantly [34]. This marks the first break in periodicity observed in the actinide atoms [24, 25]. To understand the first [24, 25] and the second [26] breaks in periodicity, we trace the evolution of the polarization-induced metastable TCS and its anionic BE as we traverse from Pu through Bk to Cf and eventually to Lr. For clarity, we refer to Table 1. The position of the R-T minimum in the polarization-induced metastable TCS is essentially fixed at roughly 0.13eV in Pu through Bk. However, the corresponding BEs of the sharp resonances in Pu, Cm and Bk TCSs are 1.22eV, 1.10eV and 0.997eV, respectively and are not located at R-T minima.

Importantly, in the TCSs of Cf the R-T minimum has flipped over to a SR appearing very close to threshold at about 0.022eV. Indeed, this marks the second break in periodicity, consistent with the observation [26]. The SR position remains essentially the same at about 0.022eV from the Cf through the Lr TCSs. But, the sharp resonances of the polarization-induced metastable TCSs are at 0.955eV, 1.03eV and 1.10eV for Cf, No and Lr, respectively. Clearly, the second break in periodicity is very sensitive to the BEs of the sharp resonances in the polarization-induced TCSs of the actinides. Notably, while the BEs associated with the polarization-induced metastable TCSs for Pu, Cm and Bk are not located at minima, those of Cf, No and Lr are, indicative of stable negative-ion formation, see Figures 2 through 4 as well as Figure 1.

The two breaks in the actinide series can also be readily understood by scrutinizing the ground states BEs of the formed negative ions, from U to Lr. Table 2 below shows that from U through Bk the BEs increase from 3.03eV to 3.55eV. But at the Pu TCSs (see Figure 2) with the ground state BE of 3.25eV the first appearance of the polarization-induced TCS (brown curve) of Pu with the BE value of 1.22eV is observed, marking the first break in periodicity of the actinides. However, from Bk to Cf the ground state BE drops from 3.55eV to 3.32eV, demonstrating the second break in periodicity. Then the BE increases from 3.32eV, the BE of Cf, through the 3.88eV BE of Lr.

**Table 2**: Periodicity breaks in the actinide series as determined from the ground states Binding Energies, BEs.

| Atom | U | Np | Pu | Cm | Bk | Cf | Es | Fm | Md | No | Lr |
|---|---|---|---|---|---|---|---|---|---|---|---|
| Z | 92 | 93 | 94 | 95 | 97 | 98 | 99 | 100 | 101 | 102 | 103 |
| BE(eV) | 3.03 | 3.06 | 3.25 | 3.32 | 3.55 | 3.32 | 3.42 | 3.47 | 3.77 | 3.83 | 3.88 |

The analysis above demonstrates the power of the Regge pole method to probe rigorously the structure and dynamics of the actinide atoms, including the confirmation of the observed first [24, 25] and second [26] breaks in periodicity in the actinide atoms without any assistance from either experiment or any other theoretical input.

### 3.7 Remarks on the Electron Affinities

It is noted here that most of the existing very sophisticated theoretical methods developed in atomic and molecular physics were tasked with reproducing experimental results, such as the EAs of multi-electron atoms with high accuracies. However, they were never intended to predict reliably the EAs; in fact, they are incapable of predicting reliably the EAs of most of the complex actinide systems, particularly those beyond the Pu atom, see Table 1. The Table 1 demonstrates that for Au, At, Eu and $C_{60}$ various theoretical calculations agree excellently with the measured EAs. These results should be used to guide the measurements/calculations of the EAs of the actinide elements. A careful examination of Table 1 reveals that the existing theoretical EAs of the actinides should be compared with those of our excited states BEs, EXT-1 and EXT-2. In particular some of the EAs have negative values and therefore are meaningless. For instance, for Pu, Bk, and Lr the absolute EA values of [27] are generally closer to our EXT-1 and EXT-2 BE values and much larger than those of [28]. For some unknown reason to us, the only available EAs of No are those of [27], but are large and negative. We surmise that many of the elaborate theoretical methods with large expansions still require to identify the vital physics underlying the EAs calculations.

### 4. Summary and conclusion

We have investigated the structure and dynamics of the actinide atoms from U to Lr through the electron elastic scattering TCSs calculated using the complex angular momentum description of scattering involving Regge poles. Dramatically sharp resonances manifesting stable negative-ion formation in the ground, metastable and excited states as well as R-T minima and SRs are found to characterize the calculated TCSs. We have exploited the sensitivity of the R-T minima and SRs characterizing the polarization-induced metastable TCSs of the U through Lr atoms to expose the first and the second breaks in periodicity in the actinide series. The deep R-T minimum near threshold first appears in the Pu TCSs, marking the first break in periodicity in the actinide elements [24, 25]. In the Cf TCSs, this R-T minimum manifests as a SR very close to threshold, demonstrating the second break [26] and continues through Lr.

Moreover, the U and the Pu TCSs exhibit significant fullerene behavior [13] through their highest excited states TCSs. This renders these atoms amenable to multi-functionalization, including nanocatalysis through their doubly-charged negative ions. And doubly-charged atomic/molecular anions have been proposed as novel dynamic tunable catalysts, with $Si^{2-}$, $Pu^{2-}$, $Pa^{2-}$ and $Sn^{2-}$ being the best catalysts; the radioactive elements usher in new application opportunities. Indeed, exploiting their radioactive nature, tunability and wide applications, ranging from water purification to biocompatibility would certainly revolutionize the field of nanocatalysis.

In conclusion, it is hoped that the rigorous results of this paper will inspire and assist both measurements and theory in the determination of the long-overdue unambiguous and reliable EAs of the actinide atoms. And the polarization-induced metastable TCSs exposed here could be useful in guiding, through the behavior of the attendant R-T minima and the SRs, the synthesis of heavier transuranic and transactinide elements.

**Table 1:** Negative-ion binding energies (BEs) and ground states Ramsauer-Townsend (R-T) minima, all in eV extracted from TCSs of the atoms and the fullerene molecule $C_{60}$. They are compared with the measured electron affinities (EAs) in eV. GRS, MS-$n$ and EXT-$n$ ($n$=1, 2) refer respectively to ground, metastable and excited states. Experimental EAs, EXPT and theoretical EAs, Theory are also included. The numbers in the square brackets are the references.

| System Z | BEs GRS | BEs MS-1 | BEs MS-2 | EAs EXPT | BEs EXT-1 | BEs EXT-2 | R-T GRS | BEs/EAs Theory | EAs GW [27] | EAs RCI [28] |
|---|---|---|---|---|---|---|---|---|---|---|
| Au 79 | 2.26 | 0.832 | - | 2.309[5] 2.301[6] 2.306[7] | 0.326 | - | 2.24 | 2.313[15] 2.50[16] 2.19[17] 2.263[56] | - | - |
| At 85 | 2.42 | 0.918 | 0.412 | 2.416[8] | 0.115 | 0.292 | 2.43 | 2.38[18] 2.42[19] 2.412[20] 2.45 [21] | - | - |
| $C_{60}$ | 2.66 | 1.86 | 1.23 | 2.684[9] 2.666[10] 2.689[11] | 0.203 | 0.378 | 2.67 | 2.663[14] 2.63[60] | - | - |
| Eu 63 | 2.63 | 1.08 | - | 0.116[1] 1.053[12] | 0.116 | - | 2.62 | 0.117[22] 0.116[23] | - | - |
| U 92 | 3.03 | 1.44 | - | 0.315[3] 0.309[4] | 0.220 | 0.507 | 3.01 | 0.175[61] 0.232[4] | 0.53 | 0.373 |
| Np 93 | 3.06 | 1.47 | - | N/A | 0.248 | 0.521 | 3.05 | - | 0.48 | 0.313 |
| Pu 94 | 3.25 | 1.57 | 1.22 | N/A | 0.225 | 0.527 | 3.24 | - | -0.276 -0.503 | 0.085 |
| Cm 96 | 3.32 | 1.57 | 1.10 | N/A | 0.258 | 0.519 | 3.31 | - | 0.283 0.449 | 0.321 |
| Bk 97 | 3.55 | 1.73 | 0.997 | N/A | 0.267 | 0.505 | 3.56 | - | -0.276 -0.503 | 0.031 |
| Cf 98 | 3.32 | 1.70 | 0.955 | N/A | 0.272 | 0.577 | 3.34 | - | -0.777 -1.013 | 0.010 0.018 |
| No 102 | 3.83 | 1.92 | 1.03 | N/A | 0.292 | 0.705 | 3.85 | - | -2.302 -2.325 | - |
| Lr 103 | 3.88 | 1.92 | 1.10 | N/A | 0.321 | 0.649 | 3.90 | 0.160[29] 0.310[29] 0.476[30] | -0.212 -0.313 | 0.295 0.465 |

**Author Contributions**
Z.F. and A.Z.M. are responsible for the conceptualization, methodology, investigation, formal analysis, and writing of the original draft. Z.L. is responsible for assembling the references and the preparation of the Tables.  A.Z.M. is also responsible for securing the funding for the research. All authors have read and agreed to the published version of the manuscript.

**Funding**
Research was supported in part by the U.S. DOE, Division of Chemical Sciences, Geosciences and Biosciences, Office of Basic Energy Sciences, Office of Energy Research, Grant: DE-FG02-97ER14743. The computing facilities of the National Energy Research Scientific Computing Center, also funded by the U.S. DOE are greatly appreciated.

**Institutional Review Board Statement**
Not applicable.

**Informed Consent Statement**
Not applicable.

**Data Availability Statement**
The original contributions presented in the study are included in the article, further inquiries can be directed to the corresponding author.

**Conflicts of Interest**
The authors declare no conflict of interest or state.